\documentclass{new_tlp}

\usepackage{fancyvrb}
\usepackage{stmaryrd}
\usepackage[noend]{algpseudocode}
\usepackage{listings} % Source Code listings.
\usepackage{graphicx} % Allows inclusion of JPEGs.
\usepackage{verbatim} % Allows begin{comment}
\usepackage{ifthen}
\usepackage{xspace}
\usepackage{amsmath}
\usepackage{amssymb}
\usepackage{amsfonts}
\usepackage{enumerate}
\usepackage{listings} 
\usepackage{tikz}
\usepackage{pstricks,pst-node}
\usepackage{url}
\usepackage{mathpartir}
\usepackage{wrapfig}

\let\oldmarginpar\marginpar
\renewcommand\marginpar[1]{\-\oldmarginpar[\raggedleft\footnotesize #1]%
    {\raggedright\footnotesize #1}}
\newcommand{\XXX}[1]{\marginpar{\textbf{XXX} #1}}
\renewcommand{\XXX}[1]{}

\newcommand{\ignore}[1]{}
\newcommand{\authornote}[3]{
  {\fbox{\sc #1}:$\blacktriangleright$\textcolor{#2}{\small{#3}}$\blacktriangleleft$}
}

\renewcommand{\authornote}[3]{}

\newcommand{\abbrevfullstop}[1]{%
  \ifthenelse{\equal{#1}{.}}{.}{%
    \ifthenelse{\equal{#1}{,} \OR \equal{#1}{;} \OR \equal{#1}{'} %
            \OR \equal{#1}{?} \OR \equal{#1}{!} %
            \OR \equal{#1}{)} \OR \equal{#1}{]} %
            \OR \equal{#1}{~}}{.#1}%
    {.\ #1}}}

\newcommand{\eg}[1]{\emph{e.g}\abbrevfullstop{#1}}
\newcommand{\ie}[1]{\emph{i.e}\abbrevfullstop{#1}}

\newcommand{\cfg}[5][scale=0.8]{
  \begin{tikzpicture}[yscale=-1,#1]
    \tikzstyle{terminal}=[draw,circle,minimum size=5pt,inner sep=0pt]
    \tikzstyle{stmt}=[draw,rectangle, minimum size=1pt, inner sep=3pt]
    \tikzstyle{edge}=[draw,thick,-stealth]

    \foreach \pos/\id in {#2}
      \node[terminal] (\id) at \pos {};

    \foreach \pos/\id/\lbl in {#3}
      \node[stmt] (\id) at \pos {\lbl};

    \foreach \start/\dest in {#4}
      \path[edge] (\start) -- (\dest);

    #5
  \end{tikzpicture}
}

\newcommand{\wtab}[2]{
  \begin{tabular*}{#2cm}{l}
    #1
  \end{tabular*}
}

\newcommand{\Prog}{\ensuremath{\mathit{Prog}}}
\newcommand{\Proc}{\ensuremath{\mathit{Proc}}}
\newcommand{\Clause}{\ensuremath{\mathit{Clause}}}
\newcommand{\Head}{\ensuremath{\mathit{Head}}}

\newcommand{\Goal}{\ensuremath{\mathit{Goal}}}
\newcommand{\Test}{\ensuremath{\mathit{Test}}}
\newcommand{\PrimTest}{\ensuremath{\olessthan}}
\newcommand{\PrimOp}{\ensuremath{\odot}}
\newcommand{\Name}{\ensuremath{\mathit{Name}}}
\newcommand{\Var}{\ensuremath{\mathit{Var}}}
\newcommand{\Val}{\ensuremath{\mathit{Val}}}
\newcommand{\Primval}{\ensuremath{\mathit{Primval}}}

\newcommand{\Func}{\ensuremath{\mathit{Func}}}
\newcommand{\Block}{\ensuremath{\mathit{Block}}}
\newcommand{\BlockExit}{\ensuremath{\mathit{BlockExit}}}
\newcommand{\Prim}{\ensuremath{\mathit{Prim}}}
\newcommand{\Phinode}{\ensuremath{\mathit{Phi}}}

\newcommand{\BlockID}{\ensuremath{\mathit{BlockID}}}
\newcommand{\return}{\ensuremath{\mathbf{return\;}}}
\newcommand{\iftest}{\ensuremath{\mathbf{if\;}}}
\newcommand{\thentest}{}
\newcommand{\elsetest}{}
\newcommand{\goto}{\ensuremath{\mathbf{goto\;}}}

\newcommand{\Const}{\ensuremath{\mathit{Const}}}

\newcommand{\lfp}{\ensuremath{\mathit{lfp}}}

\newcommand{\id}{\ensuremath{\mathsf{id}}}
\newcommand{\ret}{\ensuremath{\mathsf{ret}}}
\newcommand{\st}{\ensuremath{\mathsf{st}}}
\newcommand{\tr}{\ensuremath{\mathit{true}}}

\newcommand{\blocklbl}[1]{\ensuremath{\mathsf{#1}}}

\newcommand{\defs}{\ensuremath{\mathsf{defs}}}
\newcommand{\vars}{\ensuremath{\mathsf{vars}}}
\newcommand{\newvar}{\ensuremath{\mathsf{newvar}}}
\newcommand{\newproc}{\ensuremath{\mathsf{newproc}}}

\renewcommand{\phi}{\varphi}

\newcommand{\tuple}[1]{\ensuremath{\langle #1 \rangle}}
\newcommand{\powerset}{\ensuremath{\mathcal{P}}}
\renewcommand{\implies}{\rightarrow}
\renewcommand{\to}{\longrightarrow}

\title[Relational IR]
	{Horn Clauses as an Intermediate Representation \\ 
	for Program Analysis and Transformation%
	\thanks{This work was supported by the Australian Research 
	Council through Discovery Project Grant DP140102194.} }

\author[G. Gange et al.]
       {GRAEME GANGE\\
        Department of Computing and Information Systems \\
	The University of Melbourne, Victoria 3010, Australia \\
	\email{gkgange@unimelb.edu.au}
   \and	JORGE A. NAVAS\\
 	NASA Ames Research Center, Moffet Field CA \\
	\email{jorge.a.navaslaserna@nasa.gov}
   \and PETER SCHACHTE, 
        HARALD S{\O}NDERGAARD, PETER J. STUCKEY\\
        Department of Computing and Information Systems \\
        The University of Melbourne, Victoria 3010, Australia \\
 	\email{\{schachte,harald,pstuckey\}@unimelb.edu.au} 
       }
\date{}

\begin{document}
\maketitle

\begin{abstract}
  Many recent analyses for conventional imperative programs begin by
  transforming programs into logic programs, capitalising
  on existing LP analyses and simple LP semantics.
  We propose using logic programs as an intermediate program
  representation throughout the compilation process.
  With restrictions ensuring determinism and single-modedness,
  a logic program can easily be transformed to machine language or
  other low-level language, while maintaining the simple semantics
  that makes it suitable as a language for program analysis and
  transformation.
  We present a simple LP language that enforces determinism and
  single-modedness, and show that it makes a convenient program
  representation for analysis and transformation.
\end{abstract}


\section{Introduction}
\label{sec-intro}

Most compilers, regardless of the programming language(s) and paradigms 
supported, use some \emph{Intermediate Representation} (IR)
between parsing the input program and emitting the object code.
Use of an IR has the significant advantage of allowing a compiler to
target multiple CPU architectures, and even multiple programming
languages, without duplicating the bulk of the compiler, which
operates exclusively on the IR.
Over the course of the compilation, this representation will be
analysed for different characteristics and transformed in various
semantics-preserving ways, in preparation for efficient object
code generation.
Thus it is important for an IR to make program analysis and
transformation as simple and convenient as possible.
\emph{Three-address code} has been a popular form for this purpose
for many years.
Figure~\ref{fig:three-address-syntax} presents a three-address code
language.
Here we assume we are given 
\Name, the set of all possible function names;
\Var, the set of variable names;
and \Const, the set of all primitive constant values.
We let $\Primval = \Var \cup \Const$.
To simplify exposition, we let
$\PrimOp$ stand for all primitive arithmetic and logical operators, and
$\PrimTest$ stand for all primitive binary comparison operators.

Each \emph{basic block} of a function 
is a sequence of function calls and primitive
instructions, ending with a control transfer to another basic block.
Once control enters a basic block, it is guaranteed to
reach its end (unless some exceptional circumstance arises).
This guarantee makes analysis of each basic block straightforward.

\begin{figure}
  \begin{minipage}[t]{0.55\linewidth}
    \[
    \begin{array}{lrl}
      \Prog & \rightarrow &
                            \Func^*

      \\ \Func & \rightarrow &
                               \Head\; \Block\; \Block^*

      \\ \Block & \rightarrow &
                                \BlockID: \Prim^* \BlockExit

      \\ \BlockExit & \rightarrow &
                                    \return \Val
      \\ & \vert &
                   \iftest \Test \;
                   % \Sigmanode^* \;
                   \thentest \BlockID\; \elsetest \BlockID
      \\ & \vert &
                   \goto \BlockID

    \end{array}
    \]
  \end{minipage}
  \quad
  \begin{minipage}[t]{0.30\linewidth}
    \[
    \begin{array}{lrl}
      \Head & \rightarrow &
                               \Name(\Var^*)

      \\ \Prim & \rightarrow &
                               \Var = \Val
      \\ & \vert &
                   \Var = \Val \PrimOp \Val
      \\ & \vert &
                   \Name(\Val^*)

      \\ \Test & \rightarrow &
                               \Val \PrimTest \Val

      \\ \Val & \rightarrow &
                              \Var\; \vert\; \Const
    \end{array}
    \]
  \end{minipage}
\caption{A three-address code language}
\label{fig:three-address-syntax}
\end{figure}

A popular variant of three-address code is \emph{Static
Single Assignment} (SSA) 
form~\cite{Alpern_POPL88,Cytron_Toplas91,Lat-Adv:CGO04}.
SSA was proposed as a way to generalize \emph{value numbering}, a
technique used to remove redundant computation.
In SSA form, each variable is assigned at most once
in its scope. % (\emph{i.e.}, the whole function).
Where a variable would be reassigned, a new variable is instead 
introduced.
Since each variable is only assigned once, it is not necessary to
consider the program point when referring to a variable, only the
function it appears in.
This makes many analyses simpler and more efficient, because a single
abstract value can be associated with each variable name in a
function, and the set of variable names of interest is limited and
easily determined.

\begin{figure}[t]
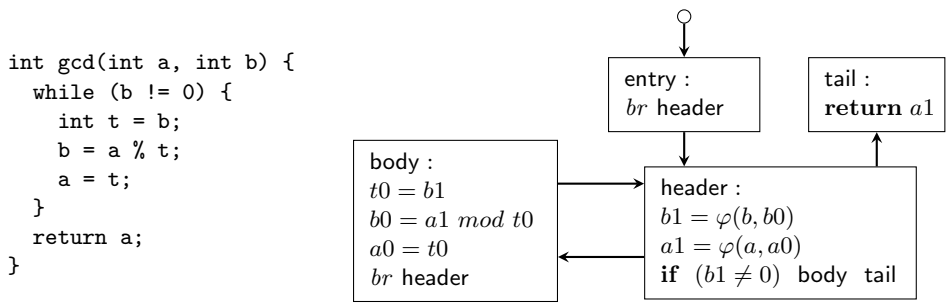

  \begin{center}
   \begin{minipage}[b]{.3\textwidth}
\begin{verbatim}
int gcd(int a, int b) {
  while (b != 0) {
    int t = b;
    b = a % t;
    a = t;
  }
  return a;
}

\end{verbatim}
\end{minipage}
\quad
    \cfg{{(0.5,-0.5)/p0}}
        {{(0.5, 0.8)/entry/\wtab{$\blocklbl{entry:}$ \\
                               $\mathit{br}$ $\blocklbl{header}$}{1.8}},
        {(2.1, 3.1)/header/\wtab{$\blocklbl{header:}$ \\
                              $b1 = \phi(b, b0)$ \\  
                              $a1 = \phi(a, a0)$ \\  
                              $\iftest$ $(b1 \neq 0)$ $\thentest$ $\blocklbl{body}$ $\elsetest$\ $\blocklbl{tail}$}{3.4}},
        {(-3.3, 2.9)/body/\wtab{$\blocklbl{body:}$ \\
                              $t0 = b1$ \\
                              $b0 = a1\ mod \ t0$ \\
                              $a0 = t0$ \\
                              $\mathit{br}$ $\blocklbl{header}$}{2.5}},
        {(3.7,0.8)/tail/\wtab{$\blocklbl{tail:}$ \\
                              $\return a1$}{1.6}}} 
      {{p0/entry}}
      {
        \path[edge] (entry) -- (header.north -| entry);
        \path[edge] (header.north -| tail) -- (tail);
        \path[edge] (header.190) -- (header.190 -| body.east);
        \path[edge] (body.20) -- (body.20 -| header.west);   
      }
  \end{center}
  \caption{The \texttt{gcd} function in C (left) and LLVM-style SSA form (right)\label{fig:gcd-ssa}
}
\end{figure}

A basic block with multiple predecessors presents a complication
for SSA: a variable use in such a block may refer to definitions of
those variables in any of the predecessor blocks.
To give such a variable a single definition, SSA introduces the concept
of a \emph{$\phi$ node}: the variable is assigned the result of a ``fake''
function that takes as input the version of the variable from each
predecessor block.
A block with multiple predecessors will contain as many $\phi$ nodes
as it has variables with alternative definitions in earlier blocks.
Figure~\ref{fig:ssa-syntax} presents the changes to three-address
syntax needed to transform to SSA:  each block may begin with $\phi$
nodes.
Consider, for example, the C code to compute the greatest
common divisor shown in Figure~\ref{fig:gcd-ssa} (left side). 
This code can be converted into SSA form as shown in
Figure~\ref{fig:gcd-ssa} (right). 

\begin{figure}
\[
\begin{array}{lrl}
  \Block & \rightarrow &
        \BlockID: \Phinode^* \Prim^* \BlockExit

\\ \Phinode & \rightarrow &
        \Var = \varphi(\Var^*)
\end{array}
\]
\caption{Changes to three-address language to produce SSA}
  \label{fig:ssa-syntax}
\end{figure}

Several researchers have presented program analyses that work by 
first transforming an imperative source program 
(\eg, \citeN{Spoto_Toplas10} and \citeN{Albert_TCS12}),
or Java bytecode (\eg, \citeN{Benton_PPDP07})
into an abstract form based on the constructs of logic
programming, and then analysing this result.
Others (\eg, \citeN{Whaley_PLS05}) have used logic programs to 
represent program analyses.
In some cases this benefits from existing logic program analyses, but
the greater benefit derives from the simple, traditional $T_P$ semantics
for logic programs, and hence simpler and more powerful analyses.
Logic programs have none of the limitations of SSA form that we detail 
below.

In this paper we propose representing
an imperative source program
as a logic program throughout the compilation process.
It may be surprising to think of compiling C programs by translation
to Prolog, rather than the reverse, but we
show that placing a few limitations on the generated logic programs
leaves low-level programs suitable for high-level analysis
and transformation, and also for final translation to machine language.

In Section~\ref{sec_problems} we discuss problematic aspects of SSA
and related forms, together with suggested ways of addressing the problem.
In Section~\ref{sec_lpvm} we introduce ``Logic Programming (LP) Form'' 
and we show how to translate a three-address code to it.
In Sections~\ref{sec-analysis} and \ref{sec-neededness} we give
example analyses for LP form.
In Section~\ref{sec-related} we discuss related work.
Finally, 
Section~\ref{sec_conclusion} reviews what has been achieved with
the proposed LP form, and concludes.

\section{SSA and Allied Forms: Problems and Solutions}
\label{sec_problems}

\begin{wrapfigure}[16]{r}{.36\textwidth}
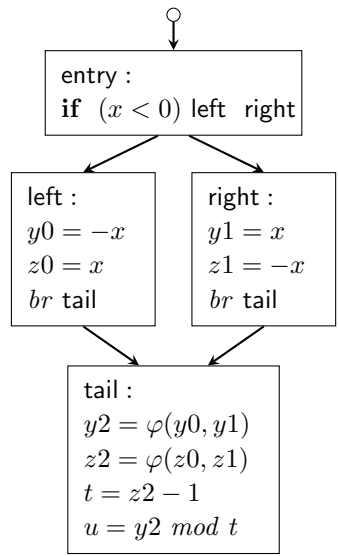

  \begin{center}
  \vspace*{-5ex}
    \cfg{{(0.0,-0.4)/p0}}
        {{(0.0, 0.9)/entry/\wtab{$\blocklbl{entry:}$ \\
          $\iftest\ (x<0)\ \thentest \blocklbl{left}\ \elsetest\ \blocklbl{right}$}{3.2}},
        {(-1.5, 3.5)/left/\wtab{$\blocklbl{left:}$ \\
          $y0 = -x$ \\  
          $z0 = x$ \\  
          $\mathit{br}$ $\blocklbl{tail}$}{1.7}},
        {(1.5, 3.5)/right/\wtab{$\blocklbl{right:}$ \\
          $y1 = x$ \\  
          $z1 = -x$ \\  
          $\mathit{br}$ $\blocklbl{tail}$}{1.7}},
        {(0.0, 7.0)/tail/\wtab{$\blocklbl{tail:}$ \\
          $y2 = \phi(y0,y1)$ \\
          $z2 = \phi(z0,z1)$ \\
          $t = z2 - 1$ \\
          $u = y2\ \mathit{mod}\ t$}{2.6}}}
            {{p0/entry}}
      {
        \path[edge] (entry.250) -- (left.north);
        \path[edge] (entry.290) -- (right.north);
        \path[edge] (left.south) -- (tail.110);
        \path[edge] (right.south) -- (tail.70);
      }
  \end{center}
  \caption{SSA and branching\label{fig:branching-ssa}
}
\end{wrapfigure}
While SSA form does simplify a number of common program analyses,
it has significant limitations that interfere with others.
Most of these problems can be solved, at the cost of further
complicating the SSA form.
In this section we will consider these limitations.

\subsection{Path obliviousness}

Basic blocks do not indicate the constraints that must be satisfied for
them to be entered.
These constraints appear in predecessor blocks.
In a forward analysis, this means constraints must be propagated from
conditional branches to their target blocks.
A backward analysis is clumsier:  it must peek backward into each
predecessor block to see what conditions hold.

Consider, for example, forward interval analysis of the code shown in
Figure~\ref{fig:branching-ssa}.
The blocks \textsf{left} and \textsf{right} both refer to ``$x$'' so
there is no straightforward way to separate the reasoning that 
needs to be done under different assumptions about $x$.
The next section considers the use of \emph{different} names for $x$ in
the separate branches, which would help in this example.
However, as it stands, we cannot assign non-trivial intervals to
$y2$ and $z2$ in the absence of path constraints that record how
control reached \textsf{tail}.

To solve the path obliviousness problem, 
\citeN{Ballance_PLDI90} have proposed the
use of \emph{Gated Single-Assignment} form (GSA).
SSA's $\phi$ nodes are replaced by different types of
\emph{gating functions}.
These capture the control conditions that determine which of the
various definitions that reach the node should provide its value.
One gating function, $\gamma$, is in essence an if-then-else function.
For example we might translate $\phi(x_1,x_2)$ to
$\gamma(P, x_1, x_2)$, where $P$ is some branch condition from
elsewhere in the program.
Flow of definitions inside loops are managed by additional
gating functions to handle initial and loop-carried values
($\mu$ nodes) as well as loop-exiting values ($\eta$ nodes).

This form makes information flow more transparent, but it is extremely
complex, compared to SSA.
The form we propose has greater uniformity, as it does not
introduce
a variety of different mechanisms for the ``joining'' or ``merging''
of information.
Moreover, GSA, as SSA, does not readily lend itself to backward analysis,
as discussed next.

\subsection{Forward bias}
The $\phi$ nodes of SSA are convenient when analysing each basic
block, as they clearly indicate which variables of which other blocks
provide values to the variables of the block.
However, this assumes \emph{forward} analysis.
In this direction, where execution paths join, each variable with
alternative sources is indicated by a $\phi$ node specifying the
different names for each alternative, and
because each variable is only defined once per function, it naturally
receives only one abstract value during forward analysis.

For a backward analysis, however,
there is no node dual to a $\phi$ to
indicate the alternative destinations that may use each variable
following a branch instruction.
(While the branch indicates alternative destinations, it does not
specify the variables that may be used there).
Importantly,
the alternative destinations for a branch all have the same name for
each variable.
In a backward analysis, then,
different blocks of a function may determine different abstract values
for the same variable name:
the virtue of SSA that each variable has a unique definition in each
function does not apply to backward analysis.

Consider Figure~\ref{fig:branching-ssa} and
suppose we wish to verify whether the division is safe, 
\ie that $t$ cannot be 0.
In a fixed-width integer context (as we assume here),
it is convenient to use ``wrapped intervals''~\cite{Gange:TOPLAS15}
as an abstract domain, as these allow us to capture both intervals
and complemented intervals.
Reasoning backwards, we find the following sufficient safety condition
for the \textsf{tail} block:
$z0, z1 \not\in [1,1]$.
For $x$, this then translates to 
$x \not\in [1,1]$
(for \textsf{left}), and 
$x \not\in [-1,-1]$
(for \textsf{right}).
This allows us to conclude that all will be well if $x \not\in [-1,1]$,
but that is insufficient to prove safety.

To address this problem, \citeN{Ananian_masters99} has proposed
adding $\sigma$ nodes to SSA form to create
\emph{Static Single Information} (SSI) form.
Where SSA form has a $\phi$ node at the top of each block indicating
where the value of each variable in the block comes from, SSI adds a
$\sigma$ node at the bottom of each branching block indicating where 
each variable's value goes to.
This permits reasoning in both directions and provides
(variable) names for all relevant pieces of information.
However, it does not address a number of other problems,
as we now explain.

\subsection{Lack of $\phi$ node compositionality}
\label{sec:compositionality}
For a non-relational \emph{value analysis}, which assigns each variable 
a single abstract value,
a $\phi$ node conveniently specifies that the abstract value for a
variable is the join of the abstract values of the input variables.
For a relational analysis, however, a $\phi$ node does not have such a
simple interpretation.
Consider an octagon analysis~\cite{Mine_Octagons_HOSC06}
of the program snippet in Figure~\ref{fig:branching-ssa}.
For the two transitions to the \textsf{tail} block, we have
$x - z0 = 0 \land y0 + z0 = 0$ and 
$x - y1 = 0 \land y1 + z1 = 0$.
Or, assuming SSI, we have
$x0 \geq 0 \land x0 - z0 = 0 \land y0 + z0 = 0$ and 
$x1 < 0 \land x1 - y1 = 0 \land y1 + z1 = 0$.
In either case, there is no meaningful (abstract)
interpretation of the statement $y2 = \phi(y0, y1)$
in isolation that does not throw away most of these relationships.
In particular, we lose the fact that $y2 + z2 = 0$.
The two $\phi$ nodes must be treated together to prevent
this loss of precision.
What is really needed is a single node that conveys the information of
$(y2,z2) = \phi((y0,z0), (y1,z1))$.

SSI does not help with this problem and in fact the remedies
discussed so far appear to address particular symptoms rather than
a more fundamental cause which, in our view, is an insufficiently
abstract view of name management.

\subsection{Name management}
The $\phi$ and $\sigma$ nodes of SSA and SSI form require special
treatment during analysis.
A $\phi$ node $v = \phi(v1,v2,\ldots)$, 
cannot be treated like a function call, because the variables mentioned
come from alternative blocks---that is, they cannot exist at the same time.
For example, when analysing a basic block beginning with
$v6 = \phi(v3,v5)$, we must find the analysis results for the two
predecessor blocks, rename $v3$ to $v6$ in the first and $v5$ to $v6$
in the second, and then find the join of the two and project away
other variables in the originating blocks.
In essence, all $\phi$ (and $\sigma$) nodes in a block must
be treated similarly to the way a function call is treated:
information about actual parameters must be renamed to match formal
parameters (or vice-versa for backward analysis), information about
variables not conveyed in the call must
be projected away, and the join of all incoming calls must be taken.

\citeN{Appel_CC92} and \citeN{Kelsey_SN95} observed similarities 
between SSA and 
continuation-passing style in functional programming.
Later \citeN{Appel_SN98} observed that SSA is in a sense equivalent to
functional programming \emph{without} continuations, and he presented
a transformation from SSA to functional program (FP) form.
This form mitigates the name management problem, using parameter 
passing to serve the purpose of $\phi$ nodes:
where SSA form would have a block with a $\phi$ node for each variable
defined in predecessor blocks, the FP form has a function with a
parameter for each variable defined outside.
Likewise, FP form uses function calls in place of jumps between blocks.
Since SSA form supports function call and return in addition to $\phi$
nodes and jumps between blocks, FP form is notably simpler than SSA.
So while analyses for SSA form are often only intra-procedural,
analyses for FP form will naturally be inter-procedural as well.

Appel's note ``SSA is functional programming''
\cite{Appel_SN98} conveys these points very clearly.
But a corollary is that functional form also preserves forward bias.
We share the enthusiasm for a declarative formalism but we also
point out that a relational view can offer
greater flexibility than a functional view.

\subsection{Input/output asymmetry}
While each input parameter of a function has a unique name apparent in
the function header,
the return value cannot be determined without scanning all the
function blocks.
In fact, there may be many alternative variables returned by different
blocks.
This is inconvenient for any summarising analysis, which ultimately
must project the analysis
result for the function onto the function inputs and output.
For example, in Figure~\ref{fig:gcd-ssa}, one knows that $a1$ and $b1$
are input to the function (this is omitted from the figure to save space),
but must examine all the blocks of the function to see that $a1$ is the 
output.
In fact, a function with more than one return may have many different 
output variables.

These problems can be avoided by first transforming the function
replacing all return statements with jumps to a distinguished new
final block containing a $\phi$ node joining all return values into a
new variable, which is then returned.
If the function header is augmented to record this final variable name
along with the function parameters, it would not be necessary to scan
all blocks to find the unique return variable.
This extra step is not difficult, but is unnecessary for LP form.

A related inconvenience is the fact that functions can only return a
single result.
If, for example, two functions compute different values through
similar computations, and the two are often called together,
it may be desirable to fuse the two functions into a single one that
returns two values.
Of course, this may be done by returning a tuple, but in this case a
structure is returned instead of two separate values, which may thwart
many analyses.
This can be solved by allowing functions to return multiple
separate values.

\subsection{Implicit variable scoping}
While the $\phi$ nodes of a block indicate some of the defined 
variables on entry to the block, they do not indicate all of them.
In fact, a block with only one predecessor will generally not have any
$\phi$ nodes at all, and so no indication at all of which variables are
defined on entry.
Neither does SSA form provide any indication of which variables of one 
block are communicated to its successors.
For analyses whose efficiency depend on minimising the number of variables
under consideration, knowing which variables enter and leave a block would
allow irrelevant variables to be projected away.

\section{LP Form}
\label{sec_lpvm}

SSA is a small refinement of three-address code.
We argue that a larger refinement, to
a restricted form of logic programming, provides the
single-assignment benefits of SSA for ease of analysis while 
avoiding the problems outlined in Section~\ref{sec_problems}.

\begin{figure}
  \begin{minipage}[t]{0.4\linewidth}
    \[
    \begin{array}{lrl}
      \Prog & \rightarrow &
                            \Proc^*

      \\ \Proc & \rightarrow &
                               \Clause^*

      \\ \Clause & \rightarrow &
                                 \Head \leftarrow \Goal^*

      \\ \Head & \rightarrow &
                               \Name (\Var^*;\Var^*)
    \end{array}
    \]
  \end{minipage}
  \quad
  \begin{minipage}[t]{0.4\linewidth}
    \[
    \begin{array}{lrl}
      \Goal & \rightarrow &
                   \PrimOp(\Val^*;\Var^*) 
      \\ & \vert &
                   \PrimTest(\Val, \Val;)
      \\ & \vert &
                   \Name(\Val^*;\Var^*) 

      \\ \Val & \rightarrow &
                              \Var\; \vert\; \Const
    \end{array}
    \]
  \end{minipage}
\caption{An LP form language\label{fig:lpvm-syntax}}
\end{figure}
\pagebreak
Figure~\ref{fig:lpvm-syntax} presents a restricted
Logic Programming language suitable for representing low-level programs.%
\footnote{Details such as handling of type information and symbol
  tables are outside the scope of this paper.  Our handling of them is
  similar to that of other IRs.}
In addition to fitting this grammar, LP form requires that for each
guard (\ie, \PrimTest) in each clause, 
there must be at least one other clause for the same procedure that is
identical up to that guard, followed by the complementary guard,
and any two clauses for a procedure must contain complementary guards,
up to which they are identical.\footnote{Note that, in LP form 
  constructed directly from three-address code, there will be at 
  most one guard in a clause;
  however, inlining can produce clauses with multiple guards.}
Furthermore, all clause heads for a given procedure must be identical.
This tames the nondeterminism of logic programming, ensuring
that exactly one clause will succeed for each set of
inputs, and makes analysis of procedures with multiple clauses simpler.
That is, only one clause of each procedure will be executed, and no
backtracking will be necessary.

This form also tames the multiple modes of a logic program by
explicitly dividing the arguments into inputs followed by outputs,
separated by a semicolon.
In calls to primitive as well as user defined procedures, 
an input argument must be either a variable or a constant
value, and an output argument must be a variable.
All parameters in procedure heads must be variables.
This ensures that variables are free until they are assigned, 
after which they are ground.
As in Mercury~\cite{Mercury}, no dereferencing is ever needed.

LP form differs from SSA form in the following ways:
\begin{itemize}
\item Instead of blocks, LP form has \emph{clauses}; a procedure
  comprises one or more clauses, exactly one of which will be executed.
\item Instead of conditional constructs and computed jumps, LP form
  has \emph{guards}, instructions that
  can either succeed or fail, determining which clause will be executed.
\item It replaces unconditional branches with procedure calls, and
  loops with recursion.
\item All registers (variables) in a clause are either parameters to
  that procedure or are defined in that clause, thus it has no need
  for $\phi$ nodes.
\item It uses parameters to pass data out of, as well as into,
  procedures, thus it has no return instruction.
\item It explicitly models changes to data structures and input/output
  operations, allowing pure functions to be recognised and optimised.
  SSA could do this, but, at least in the LLVM implementation, does not.
\item Where SSA form has four different control transfer operations, plus
  $\phi$ nodes, LP form has only procedure calls and multiple clauses, so
  LP form is simpler.
\end{itemize}
One disadvantage of this representation is that the common initial 
parts of the clauses are 
duplicated for each clause, leading to duplicated analysis effort.
Our current preliminary implementation factors out the duplicated code,
representing a procedure body as a tree, with a sequence of goals at each
node, and optionally a guard and two child nodes.
This not only avoids duplicated analysis work, but also ensures that the
clauses remain mutually exclusive and exhaustive through any program
transformations.

\subsection{Translation to LP form}

To simplify exposition, we assume the source program is presented in
three-address code form.\footnote{
Because variables in LP form are scoped to a single clause, rather than to
all the blocks of a function body, translation from SSA is actually
\emph{less} convenient than from three-address code.}
We denote by $\overline{v}$ a sequence of the 0 or more variables
comprising the set $v$.

To track side-effects, our translation uses the distinguished variable
$\st$ to denote the state of the computation, including the heap and
input/output state.
This ensures operations that may have side-effects will be executed in
the correct order, while allowing pure operations to be reordered.
We also use $\ret$ to hold the value returned by the function.%

\begin{figure}
\begin{mathpar}
    \inferrule
    {
      v = \vars(B_0,..B_n) \\
      \tuple{B_0,\overline{v},\id} \Rightarrow \tuple{B_0',C_0,\theta_0} \\ \cdots \\
      \tuple{B_n,\overline{v},\id} \Rightarrow \tuple{B_n',C_n,\theta_n} \\
      H = f(\overline{p},\st;\ret,\st \theta_0) \\
      H_0 = f_{B_0}(\overline{v},\st;\ret,\st \theta_0)) \\ \cdots \\
      H_n = f_{B_n}(\overline{v},\st;\ret,\st \theta_n))
    }
    {
      f(\overline{p}) B_0, \ldots B_n \ \Longrightarrow \ 
                    (H \leftarrow H_0) \wedge
                    (H_0 \leftarrow B_0') \wedge C_0 \wedge
                    \cdots \wedge
                    (H_n \leftarrow B_n') \wedge C_n
    } \\
%%%%%%%%%%%%%%%%%
    \inferrule{
    \tuple{\Pi,\overline{v},\theta} \Rightarrow \tuple{\Phi,C_1,\theta'} \\
    \tuple{\Xi,\overline{v},\theta'} \Rightarrow \tuple{\Theta,C_2,\theta''}
  }{
  \tuple{\Pi; \Xi, \overline{v}, \theta} \Rightarrow
  \tuple{\Phi \land \Theta,C_1\land C_2,\theta''}
  } \qquad
%%%%%%%%%%%%%%%%%
  \inferrule{
    r \in \Primval \\
    \newvar\ v' \\
    \theta' = \theta[v\mapsto v']
  }{
  \tuple{v = r,\overline{v},\theta} \Rightarrow
  \tuple{v' = r\theta,\tr,\theta'}
  } \\
%%%%%%%%%%%%%%%%%
  \inferrule{
    \newvar\ v' \\
    \theta' = \theta[v\mapsto v'] \\
    \overline{a'} = \overline{a}\theta
  }{
  \tuple{v = \PrimOp(\overline{a}),\overline{v},\theta} \Rightarrow
  \tuple{\PrimOp(\overline{a'};v'),\tr,\theta'}
  } \qquad
%%%%%%%%%%%%%%%%%
  \inferrule{
    \newvar\ v',\st' \\
    \theta' = \theta[v\mapsto v',\st\mapsto\st'] \\
    \overline{a'} = \overline{a}\theta
  }{
    \tuple{v = g(\overline{a}),\overline{v},\theta} \Rightarrow
    \tuple{g(\overline{a'},\st;v',\st'),\tr,\theta'}
  } \\
%%%%%%%%%%%%%%%%%
  \inferrule{
  }{
  \tuple{\return v,\overline{v},\theta}
  \Rightarrow
    \tuple{\ret = v\theta,\tr,\theta}
  } \qquad
%%%%%%%%%%%%%%%%%
  \inferrule{
    \newvar\ \st' \\
    \theta' = \theta[\st\mapsto\st'] \\
    }{
    \tuple{\goto B,\overline{v},\theta}
    \Rightarrow
    \tuple{f_{B}(\overline{v}, \st; \ret, \st'),\tr, \theta'}
  } \\
%%%%%%%%%%%%%%%%%
  \inferrule{
    \textsf{newproc}\ \nu \\
    C_t = \nu(\overline{v}, \st; \ret, \st') \leftarrow 
    v_i \PrimTest v_j \land
    f_{B_t}(\overline{v}, \st; \ret, \st') \\
    C_f = \nu(\overline{v}, \st; \ret, \st') \leftarrow 
    \neg v_i \PrimTest v_j \land
    f_{B_f}(\overline{v}, \st; \ret, \st')
}{
  \tuple{\iftest (v_i\PrimTest v_j)\; B_t\; B_f,\overline{v},\theta}
  \Rightarrow
  \tuple{\nu(\overline{v}, \st; \ret, \st'), C_t \land C_f, \theta}
}
\end{mathpar}
\caption{Translation from three-address code to LP form%
\label{fig:C-translation}}
\end{figure}

Figure~\ref{fig:C-translation} presents our translation.
Here the notation $\Phi \Longrightarrow \Psi$
indicates that the function
definition $\Phi$ is transformed to the clauses $\Psi$.
In the remaining transforms,
the notation
$\tuple{\Phi,\overline{v},\theta} \Rightarrow \tuple{\Psi,C,\theta'}$
means that, in the context of substitution $\theta$ and variables
$\overline{v}$, statements $\Phi$ are
translated to goals $\Psi$, with extra clauses $C$ and resulting
substitution $\theta'$.
The substitutions are used to ensure each variable has a single 
assignment, and the extra clauses are for auxiliary predicates 
generated to implement conditionals.
We let $\newvar\ x$ and $\newproc\ x$ specify that $x$ is a
fresh variable or procedure name, respectively.

\begin{figure}[t]
  \begin{align*}
    \mathit{gcd}(a,b,\st;\ret,\st')  &\leftarrow
    \mathit{gcd_{header}}(a,b,t,\st;\ret,\st')
\\[2ex]
    \mathit{gcd_{header}}(a,b,t,\st;\ret,\st')  &\leftarrow
    \mathit{gcd_{\nu}}(a,b,t,\st;\ret,\st')
\\[2ex]
    \mathit{gcd_{\nu}}(a,b,t,\st;\ret,\st')  &\leftarrow
    b \neq 0 \land
    \mathit{gcd_{body}}(a,b,t,\st;\ret,\st')
\\
    \mathit{gcd_{\nu}}(a,b,t,\st;\ret,\st')  &\leftarrow
    b = 0 \land
    \mathit{gcd_{tail}}(a,b',t,\st;\ret,\st')
\\[2ex]
    \mathit{gcd_{body}}(a,b,t,\st;\ret,\st') &\leftarrow
    t' = b \land \mathrm{mod}(a,t';b') \land a' = t' \land
    \mathit{gcd_{header}}(a',b',t',\st;\ret,\st')
\\[2ex]
    \mathit{gcd_{tail}}(a,b,t,\st;\ret,\st) &\leftarrow
    \ret = a
  \end{align*}
\caption{The \texttt{gcd} program translated to LP form\label{fig:LP_form}
}
\end{figure}

\begin{figure}[t]
  \begin{align*}
    \mathit{gcd}(a,b;\ret)  &\leftarrow
    b \neq 0 \land
    \mathrm{mod}(a,b;b') \land
    \mathit{gcd}(b,b';\ret)
\\
    \mathit{gcd}(a,b;\ret)  &\leftarrow
    b = 0 \land
    \ret = a
  \end{align*}
\caption{The translated \texttt{gcd} program of Figure~\ref{fig:LP_form}
  after simplification}
\label{fig:LP_simplified}
\end{figure}

As indicated by the first transform,
each basic block is transformed into a single clause procedure, with one
extra clause to invoke the first.
For simplicity, each of these clauses takes all the variables appearing 
in the function, plus the state variable \st\ as inputs, and
the return value variable \ret\ and the state, as modified by the block
body, as outputs.
The final transform produces a two-clause procedure for each conditional
primitive.
Because these transforms are idempotent and non-overlapping,
confluence is assured.

Figure~\ref{fig:LP_form} shows the \texttt{gcd} function of
Figure~\ref{fig:gcd-ssa} translated to LP form.
The transformation is rather simple-minded, threading every
variable to each clause.
However, the neededness analysis described in
Section~\ref{sec-neededness} allows the removal of
unnecessary variable threading, and
a simple inlining heuristic can remove unnecessary
procedures.
Figure~\ref{fig:LP_simplified} shows the translated \texttt{gcd} program
after these transformations.

\subsection{Translation from LP form to machine language}

The Mercury project~\cite{Mercury} has demonstrated that logic programs 
can be translated to very efficient executable code by tracking 
predicate determinism at compile-time and eliminating variable
dereferencing.
LP form likewise eschews unification of ``logic variables''
and the need for dereferencing, but goes further,
eliminating nondeterminism and the need for choicepoints and a machine
register to track them.
Since LP form is designed to be suitable for any language, it does not
provide its own memory management solution, and so does not need a 
register to control memory allocation.

In fact, LP form is surprisingly close to the machine language of common
computers.
Its ability to express operations with multiple outputs better reflects 
CPU capabilities than the functional restriction imposed by common 
three-address languages.
For example, the x86 architecture's \texttt{IDIV} instruction produces 
both a quotient and a remainder in separate registers, and numerous 
instructions modify flags in addition to other registers;
these are better abstracted in LP form than in three-address code.

As mentioned above, our implementation actually factors out the common
initial part of all the clauses for a procedure.
That is, each procedure is represented as a body, which is a list of 
goals optionally ending with a test to select between two (or more) 
subsequent bodies.
This representation closely matches the structure of the code to be 
generated: some straight-line code ending with a conditional
branch to one alternative and a fall through to the other.

The end of each clause is also easily translated through last call
optimisation:
if the final operation in a clause is a procedure call, that call 
is changed to an unconditional branch to the destination.
If it is a primitive, it is followed by a return instruction.
Other than this, machine code generation for LP form is similar to
SSA or three-address code.

\section{LP form analysis and transformation}
\label{sec-analysis}

\begin{figure}
  \begin{align*}
    p(x,u) &\leftarrow
             x<0 \land negate(x,y)  \land z=x \land p_1(y,z,u) \\
    p(x,u) &\leftarrow
             x\geq0 \land y=x \land negate(x,z) \land p_1(y,z,u) \\[2ex]
    p_1(y,z,u) &\leftarrow sub(z,1,t) \land mod(y,t,u) \\
  \end{align*}
  \caption{Example of Fig~\ref{fig:branching-ssa} in LP form}
  \label{fig:lp-example}
\end{figure}

In this section we show that LP form does not share the flaws discussed 
in Section~\ref{sec_problems}, and discuss its other benefits.
Consider again the example program of Figure~\ref{fig:branching-ssa}.
After simplification through inlining of simple procedures and elimination
of unnecessary dataflow, this would be expressed in LP form as shown in
Figure~\ref{fig:lp-example}.%
\footnote{Since the definition of $p_1$ is so simple, in practice it would
  be inlined, but that would only give us stronger analysis results.}
When performing a forward interval analysis on this code, the $x<0$
condition in the first clause gives the interval $[-\infty,-1]$ for $x$,
$[1,\infty]$ for $y$, and $[-\infty,-1]$ for $z$ prior to the call to $p_1$.
For the second clause, we infer $[0,\infty]$ for $x$,
$[0,\infty]$ for $y$, and $[-\infty,0]$ for $z$.
Computing the join of the abstract states for the two calls to $p_1$, we
have $y \in [0,\infty] \wedge z \in [-\infty,0]$, so analysing $p_1$ 
gives us
$y \in [0,\infty] \wedge z \in [-\infty,0] \wedge t\in[-\infty,-1]$ on
reaching the first call to $mod$,
allowing us to certify the safety of the mod operation.
The path-awareness of LP form gives us stronger analysis results without 
any extra effort.

Since each LP form clause is logically an unordered conjunction,
it is equally adept at forward and backward analysis.
Consider a backward analysis of the program of Figure~\ref{fig:lp-example}
to determine the safety of modulo (division) operations.
This will start with the constraint $t\neq0$ at the end of $p_1$, which
implies $z\neq 1$ on entry to $p_1$.
Analysing the first clause of $p$ backwards from its call to $p_1$, 
we deduce $z\neq1 \lor x\neq1 \lor y\neq-1$ before reaching the $x<0$ goal.
Handling this goal gives us
$x<0 \implies z\neq1 \lor x\neq1 \lor y\neq-1 \equiv \mathit{True}$,
meaning we have nothing else to prove for that clause.
Turning to the second clause of $p$, we derive
$x\geq0 \implies z\neq1 \lor x\neq-1 \lor y\neq-1 \equiv \mathit{True}$,
and again the proof obligation is discharged.

Relational analyses do not present any difficulty for LP form, because it
has no artificial $\phi$ nodes to separately combine alternative versions
of variables.
This is handled through conventional procedure calls,
where the least upper bound is used to combine results for multiple calls.
Consider an octagon analysis~\cite{Mine_Octagons_HOSC06}
of Figure~\ref{fig:lp-example}.
Much like the analysis discussed in Section~\ref{sec:compositionality},
analysis derives
$y + x = 0 \land z - x = 0 \land y + z = 0$
leading to the call to $p_1$ from clause~1, and
$y - x = 0 \land x + z = 0 \land y + z = 0$
for clause~2.
Procedure calls are handled by projecting the abstract state onto the
variables appearing in the call, and computing the least upper bound of 
the states. 
In this case, this yields 
$y + z = 0 \sqcup y + z = 0 \equiv y + z = 0$,
preserving the strong results obtained for both clauses.

The other issues for SSA and FP form discussed 
in Section~\ref{sec_problems} are trivially addressed by LP form.
Lacking $\phi$ nodes, LP form has no issue with name management.
Because LP form is relational, it has no issue with input/output asymmetry.
And because each clause has its own scope, the scope of each variable is
obvious.

\section{Specialised analyses for LP form}
\label{sec-neededness}

Liveness analysis is a standard program analysis used to determine for 
each program point the set of variables whose values may be needed later.
The single assignment property enjoyed by SSA, FP, and LP forms somewhat
simplifies this analysis:  because each variable is assigned only once, 
it is not necessary to take account of variable re-assignment.
Within a single block (clause) of SSA (LP form) code, this is easily 
done by traversing the statements backward, noting the first encountered
use of each variable, which will be the last use on forward execution, 
and each variable assignment, which will be the definition of that 
variable.
To handle liveness for a whole function, analysis results must be 
propagated backward between blocks.

Dead code elimination is a transformation to remove unnecessary code.
Any code that assigns only dead variables can be removed, but doing so 
may remove variable uses, and produce stronger results for liveness 
analysis.
Thus it is beneficial to perform liveness analysis and dead code 
elimination simultaneously.
If this is extended beyond individual functions to an entire module 
or even a whole program, more dead code can be eliminated.

We present a two-phase interprocedural \emph{neededness} analysis, which
combines liveness and dead code elimination.
The first phase computes \emph{neededness dependencies}, conjunctions of
implications of the form
$x \implies y$ signifying that if variable $x$ is needed on completion 
of a goal, then $y$ is needed on entry.
This analysis can be performed bottom-up over a module's call graph, one
strongly connected component (SCC) at a time, which ensures that
all callers of a given procedure, except those in
the same SCC, will be analyzed before the procedure itself.
A fixed point must be computed for each SCC, but no iteration is necessary
between SCCs.
This reduces the number of procedures analyzed in each fixed point
iteration, since SCCs are typically fairly small.
 
Formally, we define our neededness dependency domain $N$ as the set of 
conjunctions of variable $\implies$ variable implications,
where an individual implication $x\implies y$ indicates that if 
variable $x$ is needed, then so is $y$.
We let $S$ denote the $\Goal\to N$ \emph{neededness dictionary} function 
space, specifying neededness dependencies for many procedures.
We define our analysis with the following functions:
\\[-3ex]
\begin{minipage}[t]{.45\linewidth}
  \begin{align*}
    P_d &:: \powerset(\Proc) \to S \\
    D_d &:: \Proc \to S
  \end{align*}
\end{minipage}
\begin{minipage}[t]{.45\linewidth}
  \begin{align*}
    C_d &:: \Goal^* \to S \to N \\
    G_d &:: \Goal \to \powerset(\Var) \to S \to N
  \end{align*}
\end{minipage}
\\[1ex]
Here $P_d$ gives the neededness dictionary
for all the procedures in the module;
$D_d$ yields the dictionary for a single procedure;
$C_d$ produces the neededness of a single clause given a neededness 
dictionary;
and $G_d$ gives the neededness of a single goal given the set of 
variables needed later in the clause body and a neededness dictionary.
\begin{figure}
  \begin{align*}
    P_d\ S 
    &= \lfp \left(\bigsqcup_{d\in S} D_d\ d\right) \\
    D_d\ [p(\overline{v_i};\overline{v_o}) \leftarrow (B_1, \lor\cdots\lor B_n)]\ A
    &= \lambda p(\overline{v_i};\overline{v_o})\ .\ 
      \left(\exists(\Var \setminus v_i \setminus v_o)\ .\ 
      \bigwedge_{1 \leq k \leq n} C_d\ B_k\ A\right) \\[2ex]
    C_d\ (g_1 \land \cdots \land g_n)\ A
    &= \bigwedge_{1 \leq k \leq n}
      G_d\ g_k\ \defs(g_{k+1}\land\cdots\land g_n)\ A \\[2ex]
    G_d\ \PrimOp(\overline{v_i};\overline{v_o})\ V\ A
    &= \bigwedge_{x\in v_i} \bigwedge_{y\in v_o} y \implies x \\
    G_d\ \PrimTest(x,y)\ V\ A
    &= \bigwedge_{v\in V} (v \implies x \land v \implies y) \\
    G_d\ p(\overline{v_i};\overline{v_o})\ V\ A 
    &= A\ p(\overline{v_i};\overline{v_o})
  \end{align*}
  \caption{Neededness abstract interpretation}
  \label{fig:neededness}
\end{figure}

As shown in Figure~\ref{fig:neededness},
the neededness analysis of a module is the least fixed point of the
combination of results for all procedures in the module,
and the result for a procedure is just the conjunction of the neededness 
of all its clauses, which is the conjunction of results for all
goals in each clause.
The analysis result for a primitive operation is the conjunction of
$x\implies y$ implications for each output $x$ and each input $y$.
For a primitive comparison operation, it is the conjunction of
$x\implies y$ for each variable $x$ defined later in the clause
(determined by the \defs\ function) and each input $y$ of the comparison.
Since primitive comparisons are guards, they are only needed to 
determine if the following code is executed, so they are only 
needed if some variable defined later is needed.

The second analysis phase uses these dependencies to determine which 
procedure inputs and outputs are actually used, beginning
by marking all parameters of public (exported) functions as needed.
This analysis then proceeds top-down by SCCs through the program call 
graph, with each SCC processed until a fixed point is reached.
In each iteration, each clause in the SCC is processed with a needed 
variable formula, initially the conjunction of the set of output 
variables of that procedure that are marked as needed.

Processing of a clause proceeds from last goal to first.
If any output of a goal is in the needed variable formula, the goal is
marked as needed, and the called procedure has its needed outputs marked 
for when \emph{it} is processed.
Then the neededness dictionary for the called procedure is conjoined with
the current needed variable formula, and the goal's output variables are
projected out, to produce the new needed variable formula.
This formula then comprises the live variable set for that goal.
Primitive goals are simpler:  if any output is in the needed variable
formula, it is marked as needed, its inputs are added to the
needed variable formula, and its outputs are projected out.
Once a fixed point has been reached, any goal or
parameter not marked as needed can be removed.

\section{Related Work}
\label{sec-related}

Many variants of SSA have been proposed
\cite{Ballance_PLDI90,Gerlek_Toplas95,Chow_CC96,Ananian_masters99}
and much work has been concerned with how to generate (compact)
SSA and its variants efficiently 
\cite{Cytron_Toplas91,Tu_Padua_PLDI95,Ananian_masters99}.
In Section \ref{sec_problems} we mentioned the work on FP form by
\citeN{Kelsey_SN95} and \citeN{Appel_SN98}.
\citeN{Appel_SN98} in fact sketches two translations to FP form,
one producing a ``flat'' sequence of function definitions,
the other producing nested definitions.
The latter uses fewer functions and variables and Appel points out that
the structure of function nesting makes the dominance
properties of the original control-flow graph explicit.
\citeN{Appel_SN98} also uses the equivalent of SSI's ``$\sigma$ nodes''
as a pedagogic tool; the $\sigma$ nodes are in fact pushed into 
successor blocks and become mere ``renaming'' $\varphi$ nodes.

\citeN{Peralta_Cruz_LOPSTR96} briefly
sketched a translation from SSA to CLP,
but provided no formal definition of the translation.
From their examples it is clear that the translation
differs from the one suggested here.
\citeN{Peralta_Gallagher_Saglam_SAS98}
showed how to use CLP analysis tools to analyse imperative programs.
Their approach is based on having an interpreter, written in
CLP, for the imperative language and then translating (small)
imperative programs through partial evaluation of the interpreter.

\citeN{Spoto_Toplas10} implement a termination analyzer for Java
bytecode by expressing path-length reasoning as a CLP program and
leveraging from existing CLP termination analysis tools.
The resulting analyzer is robust and entirely automatic,
covering the full language of Java bytecode.
\citeN{Albert_TCS12} use a similar approach to cost analysis.

\citeN{Morales-TPLP14} explore the use of a logic programming language
for the implementation of efficient abstract machines and runtime systems.
To this end they use a Prolog variant with certain imperative features
(mutable variables) that enables translation into efficient C-style code
while still allowing for high-level program transformations,
such as partial evaluation of instruction definitions.

CLP has been also used as the basis for software model
checking~\cite{DelzannoP99,Flanagan03} of concurrent systems
and its use in software verification tools is rapidly growing.
For example, it has been adopted in
Threader~\cite{threader}, 
UFO~\cite{ufo},
SeaHorn~\cite{seahorn-svcomp15}, 
HSF~\cite{GrebenshchikovLPR12},
VeriMAP~\cite{VeriMAP}, 
Eldarica~\cite{RummerHK13}, and
TRACER~\cite{TRACER}. 
The task of encoding verification conditions is 
different to our aim of providing a platform for program compilation, 
although both require a convenient representation for reasoning
about programs.

\section{Conclusions}
\label{sec_conclusion}

We have described Static Single Assignment form, and discussed a number 
of problems it causes for sophisticated analyses.
Many of these problems have been previously addressed, but no previous 
work has addressed all of them.
One approach that addressed several of these problems re-conceives a 
low-level program as a functional program.

We propose going further and viewing a low-level program as a logic 
program, and have suggested a simple, deterministic, strongly moded 
logic programming language as a compiler intermediate representation.
The language is fully declarative; many existing analyses for logic
programming languages will apply directly.
We have presented a powerful analysis and transformation for this form.
Because LP form uses procedure calls for all control transfer, 
operations that cross block boundaries are naturally inter-procedural.
Owing to determinism and single-mode restrictions,
LP form is surprisingly close to machine language, so final code 
generation is not difficult.
Thus LP form is a suitable choice for a compiler's intermediate code
representation.

We are currently developing an implementation of LP form, which we call
LPVM.
This is being used as intermediate representation for a compiler we are
developing for a language combining the benefits of declarative and
imperative programming.
Since the procedures of the language support multiple outputs, that 
facility in LP form is particularly important.
Rather than duplicating the extensive work of the LLVM project in 
producing high-quality, peep-hole optimised assembly language for 
multiple architectures, we plan to do all program analysis and 
transformation in LP form, and finally translate to LLVM for final 
code generation.

\bibliographystyle{acmtrans}
\bibliography{ssa}

\end{document}